\documentclass{aa}  

\usepackage{graphicx}
\usepackage{multirow}
\usepackage{txfonts}
\usepackage{xcolor}

\begin{document}

   \title{A very young $\tau$-Herculid meteor cluster observed during a 2022 shower outburst}

   \author{P. Koten \inst{1}
		  \and
        D. \v{C}apek \inst{1}
        \and 
		  J. T\'{o}th \inst{2}
		  \and
		  L. Shrben\'{y} \inst{1}
		  \and
		  J. Borovi\v{c}ka \inst{1}
		  \and
		  J. Vaubaillon \inst{3}
		  \and
		  F. Zander \inst{4}
		  \and
		  D. Buttsworth \inst{4}
		  \and
		  S. Loehle \inst{5}
          }

   \institute{Astronomical Institute, Czech Academy of Sciences, Fri\v{c}ova 298, 25165 Ond\v{r}ejov, Czech Republic\\
              \email{pavel.koten@asu.cas.cz}
              \and
              Faculty of Mathematics, Physics and Informatics, Comenius University, Mlynsk\'{a} Dolina, 842 48 Bratislava, Slovakia
              \and
              IMCCE, CNRS, Observatoire de Paris, 77 av. Denfert-Rochereau, Paris, France
              \and
              IAESS, University of Southern Queensland, West St., Toowoomba, Australia
              \and
              HEFDiG, Institute of Space Systems, University of Stuttgart, Pfaffenwaldring, Stuttgart, Germany
              }

   \date{Received dd-mm-yyyy; accepted dd-mm-yyyy}

 
  \abstract
   {To date only very few meteor clusters have been instrumentally recorded. This means that every new detection is an important contribution to the understanding of these phenomena, which are thought to be evidence of the meteoroid fragmentation in the Solar System.}
   {On 31$^{st}$ May 2022, at 6:48:55 UT, a cluster consisting of 52 meteors was detected within 8.5 seconds during a predicted outburst of the $\tau$-Herculid meteor shower. The aim of this paper is to reconstruct the atmospheric trajectories of the meteors and use the collected information to deduce the origin of the cluster.}
   {The meteors were recorded by two video cameras during an airborne campaign. Due to only the single station observation, their trajectories were estimated under the assumption that they belonged to the meteor shower. The mutual positions of the fragments, together with their photometric masses, was used to model the processes leading to the formation of the cluster.}
   {The physical properties of the cluster meteors are very similar to the properties of the $\tau$-Herculids. This finding confirms the assumption of the shower membership used for the computation of atmospheric trajectories. This was the third cluster that we have studied in detail, but the first one where we do not see the mass separation of the particles. The  cluster is probably less than 2.5 days old, which is too short for such a complete mass separation. Such an age would imply disintegration due to thermal stress. However, we cannot rule out an age of only a few hours, which would allow for other fragmentation mechanisms.}
   {}


   \titlerunning{$\tau$-Herculid meteor cluster}
   \authorrunning{P. Koten et al.}

   \maketitle
%

\section{Introduction}

The $\tau$-Herculid meteor shower\footnote{061 TAH code in IAU MDC database} is a relatively weak meteor shower, which typically displays minimal activity. Only a handful of photographic and video orbits have been reported in the past \citep{Rao2021}. It changed after the breakup of its parent body, comet 73/P~Schwassmann-Wachmann~3, which occurred in 1995 \citep{Crovisier1996}. Several authors have modelled the orbital evolution of the dust particles released during this fragmentation process and predicted a meteor outburst or even a storm on 31 May 2022 \citep[for a review see][]{Ye2022}. An outburst did indeed occur at the predicted time \citep[e.g.][]{Egal2023}.

Meteor outbursts or meteor storms are events that offer a higher chance of detecting meteor clusters. This phenomenon was observed in the case of the Leonid meteor shower storms at the turn of the century, when three such events were reported by Japanese astronomers \citep{Kinoshita1999, Watanabe2002, Watanabe2003}. Another case was a statistical grouping of Leonid meteors above chance level during maxima activity in 1999 \citep{Toth2004}. The paper of \citet{Watanabe2003} suggests that thermal stressing of very fragile cometary dust is the most likely process leading to the formation of the clusters. This was recently confirmed by \citet{Capek2022}, who analysed in detail a cluster of 10 September $\epsilon$-Perseid meteors observed over the Czech Republic in 2016 and showed that thermal stress of fragile cometary material was the most likely process leading to the formation of the cluster. They found that the age of the cluster was only about 2.3 days. Another event with the same formation scenario was a huge meteor cluster observed in northern Europe in 2022, consisting of at least 22 meteors \citep{Koten2024}. In this case, the age of the cluster was 10.6$\pm$1.7 days. It was a sporadic meteor cluster on cometary orbit. For both clusters, the ejection velocities were found to be very low, in the order of 0.1~m~s$^{-1}$.

There are no more meteor clusters that have been analysed in detail up to now. It is obvious that data on more clusters are needed to determine whether the thermal stress alone is the cause of the cluster formation, or whether other scenarios are possible. These include fragmentation due to fast rotation and collisions with other particles \citep{Capek2022}. 

Because meteor clusters are rare and unpredictable events, it is necessary to take every opportunity to increase the chances of detecting them. Predicted meteor outbursts are one of them. Although the main reason for observing the predicted $\tau$-Herculid meteor outburst was to detect the particles release from the parent comet during its disintegration in 1995, increased likelihood of the cluster detection was also one of the objectives.

This has indeed happened. A large meteor cluster was recorded by cameras on board an aircraft conducting a research flight dedicated to the observation of the $\tau$-Herculid meteor shower \citep{Vaubaillon2023}. This original paper (hereafter Paper I) reported 38 meteors, provided a basic description of the event, and estimated the probability of its occurrence. However, as the event was observed from only one station, the trajectories of the meteors were not reconstructed.

In this paper, we extend the original dataset by additional meteors found on another record obtained by different camera on the board of the same aircraft. In total, 52 meteors were identified. Their atmospheric trajectories were reconstructed under the assumption that all meteors belong to the 2022 $\tau$-Herculid meteor shower. Furthermore, the photometric masses of the meteors were calculated. All these obtained information were finally used for the investigation of probable fragmentation mechanisms of the original meteoroid.

The structure of the paper is as follows. Section \ref{instrument} describes the airborne observational campaign, the instruments used for data acquisition, and the processing procedures, including the software modification for the single station meteors. Section \ref{results} presents the results of the cluster analysis. The basic characteristics of the cluster, and the atmospheric trajectories are compared with the properties of the entire meteor shower. The masses and mutual positions of the meteors in the cluster are used for the analysis of the cluster age and ejection velocities. Finally, the origin of the meteor cluster is discussed in Section \ref{discussion}.

\section{Instrumentation and data processing}
\label{instrument}

The purpose of the airborne campaign during the $\tau$-Herculid meteor shower in 2022 was to gather valuable data on the composition, trajectory, and behaviour of the meteoroids as they entered the Earth's atmosphere.  The airborne campaign to observe the $\tau$-Herculid meteor shower was conducted on 31~May 2022 with a Phenom 300 business jet led by Rocket Technologies International (RTI, Australia) and the University of Southern Queensland (Australia). The aircraft was deployed and flew west from Dallas/Fort Worth airport, turning back over Arizona and returning to that airport. 

On board were several imaging and spectral instruments from various participating institutions. Besides the main campaign organizers RTI and The University of Southern Queensland with imaging cameras on both sides of the aircraft, there were also imaging cameras from IMCCE, Paris Observatory (France) set up on the left side of the aircraft, a spectral camera from HEFDiG (High Enthalpy Flow Diagnostics Group, Institute of Space Systems, University of Stuttgart, Germany) on the right side of the aircraft and  three spectral imaging cameras AMOS-Spec from Comenius University in Bratislava (Slovakia) on the left side of the aircraft. Each of AMOS-Spec camera was based on a CMOS DMK 33UX252 (resolution of 2048 x 1536 px and set to 14 fps) equipped with a 6 mm, F/1.4 lens, providing a FOV of 60 x 45 deg and 16 mm F/1.8 and 35 mm F/1.4. Only the recording from 6 mm lens detected the cluster and was used in this work.

The cluster was captured by two cameras -- one of the AMOS-Spec cameras and the IMCCE camera \citep{Vaubaillon2023}. Both cameras were pointed to the north at about 30 degrees above the horizon. As the AMOS-Spec was slightly shifted to the right, had a wider field of view and was slightly more sensitive, it was able to detect more meteors. This configuration allowed the detection of 52 meteors within 8.5 s. 

This cluster observation was not reported by any of the ground-based meteor networks. The nearest AMOS ground station was 780 km away.

\begin{table*}
\caption{The aircraft positions at the time of the first and the last cluster meteor.}
\label{tab_aircraft_positions}      
\centering          
\begin{tabular}{l r r r r} 
\hline
					&	Time [UT]		& 	Longitude W					&	Latitude N				&	Altitude		\\
\hline
meteor \#1		&	6:48:55.9		&	101.87100$^{\circ}$ 		&	34.19607$^{\circ}$ 		&	14.203 km		\\
meteor \#52		&	6:49:04.5		&	101.85265$^{\circ}$ 		&	34.19159$^{\circ}$ 		&	14.203 km		\\
\hline                  
\end{tabular}
\end{table*}

The aircraft positions were continuously recorded by GPS during the entire flight, with a temporal resolution of 2.8 seconds. The precise position of the aircraft (i.e. cameras) at the time of meteor appearance was then calculated by interpolation from the recorded data. The values for the time of the first and the last meteors are shown in Table~\ref{tab_aircraft_positions}. The aircraft moved by 1.76 km during this time interval.

All the meteor records were manually measured using the FishScan software \citep{Borovicka2022}. 46 of them were measured on the AMOS-Spec camera record, another 6 on the IMCCE record. The astrometric calibration for each video was done only once at midpoint of the cluster appearance. This resulted in a maximum positional deviation of 0.016 degrees at the time of both the first and the last meteor. This value is slightly smaller than the standard deviation calculated for the star positions on the calibration image. Therefore, we decided not to do positional calibration for each meteor separately. The meteor atmospheric trajectories were calculated using a newly added module to the Boltrack program \citep{Borovicka2022}, which allows the estimation of single-station meteor trajectories under the assumption of the meteor radiant and velocity. In this case the radiant and velocity of the $\tau$-Herculid meteor shower were taken into account. The exact position of the aircraft at the time of each meteor occurrence was taken into account in order to account for aircraft movement.

In the atmosphere, meteor trajectories can be rigorously determined by triangulation only if the meteor has been observed from at least two widely separated stations. Nevertheless, in case of shower meteors when the radiant and entry velocity are known, the atmospheric trajectory can be determined with sufficient accuracy even from single station video data. First, a main circle is drawn through the positional measurements. A point on the main circle closest to the shower radiant is chosen as the meteor radiant. Now, the orientation of the trajectory is fixed but the distance from the station is still unknown. That can be determined from the angular velocity of the meteor. The distance is set so that the physical velocity computed from the angular velocity corresponds to the expected meteor velocity.

In case of longer meteors, when the angular velocity changes along the path (from geometrical reasons or due to deceleration), a set of solutions can be computed for a reasonable range of assumed beginning heights of the meteor. The velocity profile along the meteor is computed for each solution. The solution with the expected beginning velocity is selected. For a better comparison with shower parameters, both the radiant and the velocity can be transformed to their geocentric values. 

As mentioned above, the atmospheric trajectories were reconstructed under the assumption that the meteors belong to the  $\tau$-Herculid meteor shower. In this case, the 2022~outburst is taken into account, not the regular meteor shower. Since the data reported by \citet{Koten2023} were recorded hours before the peak of activity, the mean geocentric radiant $\alpha_{G} = 208.6^{\circ}$, $\delta_{G} = 27.7^{\circ}$ and geocentric velocity $v_{G}$ = 11.9~km~s$^{-1}$ calculated using the Global Meteor Network data \citep{Vida2022} were used to determine the atmospheric trajectories of the single station meteors.

The absolute magnitudes of the meteors were measured at each frame and the photometric mass of each meteor was calculated \citep{Ceplecha1987}. The luminous efficiency calculated according to \citet{Pecina1983} was used for this purpose.

\section{Results}
\label{results}

\begin{figure}
  \resizebox{\hsize}{!}{\includegraphics{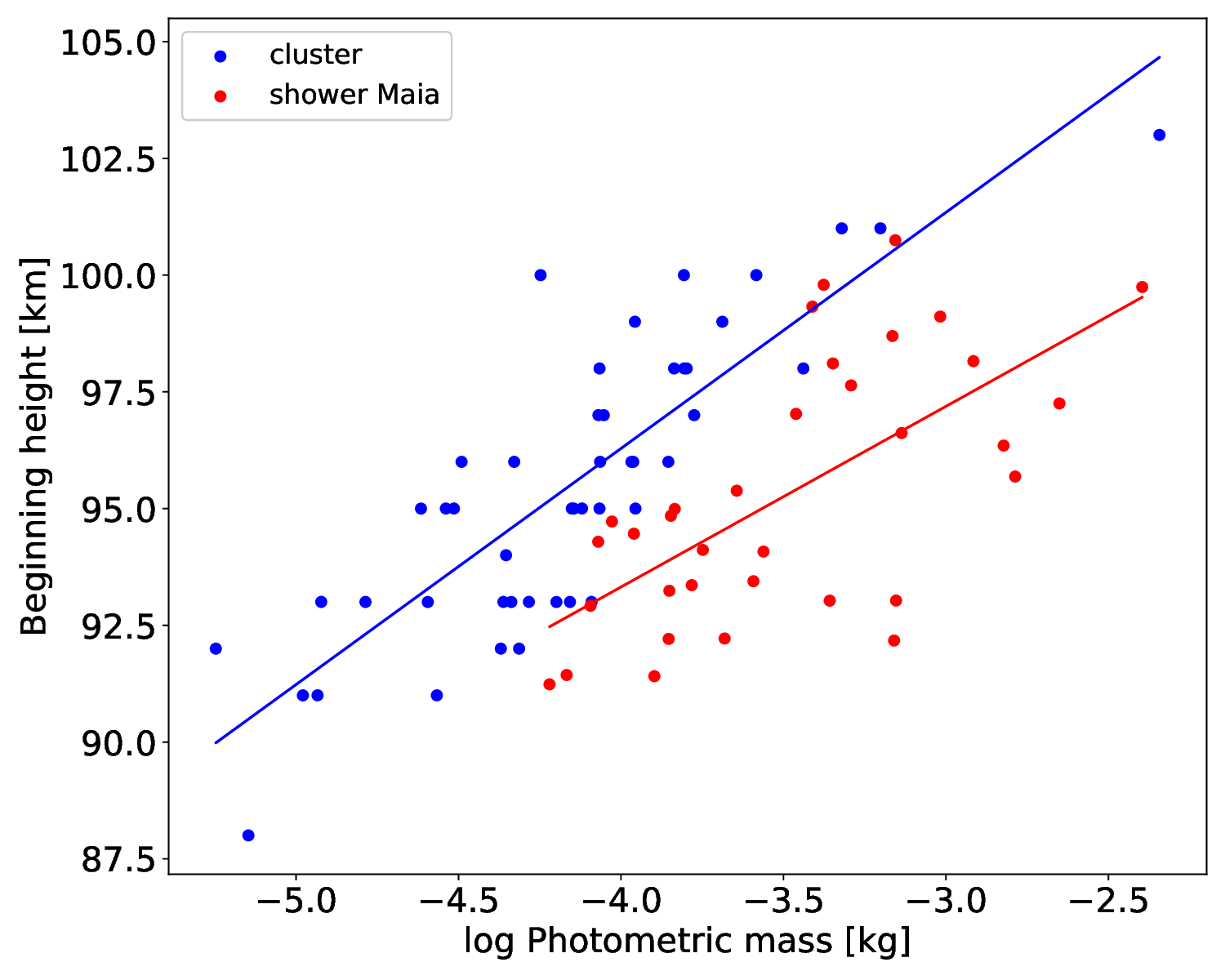}}
  \caption{Beginning height of single station cluster meteors as a function of the photometric mass (blue marks). For comparison, the beginning heights of the shower meteors recorded by the Maia cameras in the Czech Republic before the peak of the activity are added (red marks). The lower sensitivity of these cameras is responsible for the slightly lower beginning heights of the meteors.}
  \label{fig_HB}
\end{figure}

\subsection{Atmospheric trajectories}
\label{trajectories}

It is worth mentioning that the method used to estimate the trajectory of the single station meteors does not provide a unique solution. The software provides a set of meteor radiants with appropriate beginning heights and the velocities. The final choice always depends on the operator's decision. As we are looking for the $\tau$-Herculid meteors, we choose one of the radiants close to the shower radiant. Even in this case, there may be more radiants that satisfy the conditions. In this case, the plot of the beginning height against the photometric mass of the meteor is helpful (Figure~\ref{fig_HB}). From this figure, the operator can see whether the selected solution is consistent with a general trend of the beginning height within the cluster.

The correct determination of the angular velocity is crucial. Especially for very faint meteors, which are detected in only a few frames, it is difficult to measure the angular velocity correctly. In such cases it can be helpful to fix the angular velocity to the same value as that of another -- longer -- nearby meteor. A note on the angular velocity determination can be found in Appendix A.

A total of 52 meteors were found on both videos. Due to additional meteors found on the AMOS-Spec video, it was not possible to keep the same numbering as it was used in Paper I. The meteors were numbered according to the time of their appearance. The very first meteor appeared at 6:48:55.90 UT and was only recorded by the AMOS-Spec camera. Only 0.06 s later the first meteor reported in Paper I appeared. The brightest one, originally number 3, newly number 4, was detected 0.34 s later. Other meteors quickly followed within the next 8 seconds. The last one appeared at 6:49:04.45 UT.

The brightest meteor started its luminous trajectory at 103 km and ended at 89.7 km. It reached a maximum magnitude of -1.7 and its photometric mass was 4.5 grams. The meteor moved almost exactly from west to east, i.e. parallel to the direction of the aircraft, and its trajectory was relatively steep, deviating 27.4~degrees from the vertical. 

\begin{figure}
  \resizebox{\hsize}{!}{\includegraphics{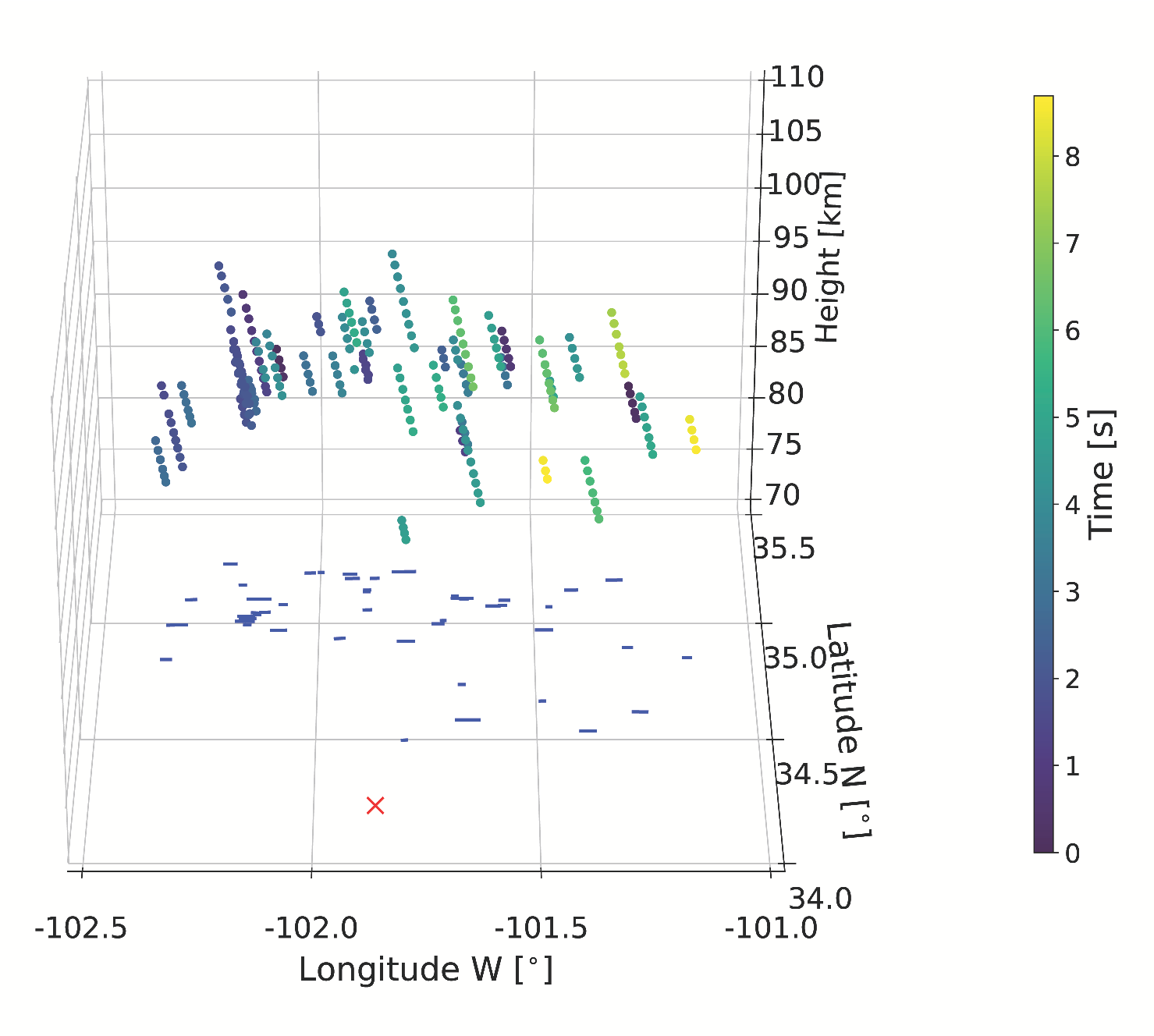}}
  \caption{3D representation of meteor atmospheric trajectories. The time from the start of the first meteor expressed in seconds is colour coded. Time T = 0 corresponds to 6:48:55.9~UT. A total of 47 single-station meteors are shown. The position of the aircraft is plotted as red cross. All meteor and the aircraft positions were projected at the altitude of 70~km.}
  \label{fig_3D_trajectories}
\end{figure}

For 5 meteors (numbers 22, 25, 29, 30, and 49) no reasonable combination of the beginning height, angular velocity, initial velocity, and radiant could be found. Among them, there were the faintest meteors, which were measured only on very few video frames. These meteors were not used for further analysis. Basic data on the atmospheric trajectories of the remaining 47 meteors are summarised in Table~\ref{tab_atm_traj}. The 3-D view of the cluster is shown in Figure~\ref{fig_3D_trajectories}. The time elapsed from the first appearance of the first meteor is colour coded. 

Although it appears that the first meteors were detected in the western part of the field-of-view and the last ones in the eastern part, this trend is not general. Some of the meteors, especially those from the beginning of the cluster, appeared more to the east than they should have according to their time of occurrence if the cluster had been perfectly aligned. 

The cluster was relatively compact. When projected onto the ground, all the trajectories covered an area of 112 x 86~km, or about 9630~km$^{2}$. It was even more compact than the September $\epsilon$-Perseid cluster, for which the projected area was 11500~km$^{2}$ \citep{Koten2017}.

Comparison of the geocentric radiants of the cluster meteors with other observations and the modelled data shows relatively good agreement between the cluster and other data. It confirms that the assumption that the cluster radiant is consistent with the shower radiant was correct. 

\begin{table*}
\caption{Atmospheric trajectories of 47 single-station meteors belonging to the $\tau$-Herculid 31 May, 2022 meteor cluster. The accuracy of the height determination is 2 km. The times represent the first appearance of each meteor.}
\label{tab_atm_traj}
\centering          
\begin{tabular}{l c r c c l l r c c c c} 
\hline
$met$	& 	$time$ 			& 	$H_{BEG}$   	&   $H_{END}$ 	&  		$H_{MAX}$  		& \multicolumn{1}{c}{$\lambda_{B}$}    & \multicolumn{1}{c}{$\phi_{B}$}     & 	\multicolumn{1}{c}{$l$}    &   	$t_{dur}$  &  	$M_{MAX}$  &  $m_{phot}$     &   $K_{B}$ 	\\
\#		&	[UT]			&	[km]			&	[km]		&		[km]	   		&	\multicolumn{1}{c}{W [$^{\circ}$]}   &   \multicolumn{1}{c}{N [$^{\circ}$]}    & [km]   &	[s]        &	[mag]	   &  [kg]           &              \\
\hline
01      &	6:48:55.90      &  	 95           	&       91.5    &       94.0    	&	-101.29535	& 	34.85618	&	 4.0   &  	0.29       &  	 3.0       &     7.2E-05   &     6.5           \\           
02      &	6:48:55.96      &  	 95           	&       91.5    &       93.5    	&	-102.09167	& 	35.04656	&	 3.9   &  	0.21       &  	 2.3       &     8.6E-05   &     6.6           \\           
03      &	6:48:56.36      &  	 96           	&       92.2    &       93.2    	&	-101.57803	& 	35.06523	&	 4.4   &  	0.29       &  	 3.1       &     4.7E-05   &     6.7           \\           
04      &	6:48:56.36      &  	103           	&       89.7    &       94.7    	&	-102.16682	& 	35.07038	&	15.0   &  	0.71       &  	-1.7       &     4.5E-03   &     6.4           \\           
05      &	6:48:56.91      &  	 95           	&       90.4    &       92.0    	&	-101.67467	& 	34.69566	&	 5.2   &  	0.14       &  	 3.4       &     2.4E-05   &     6.8           \\           
06      &	6:48:57.21      &  	 93           	&       89.0    &       91.8    	&	-101.89473	& 	35.11117	&	 4.5   &  	0.14       &  	 2.9       &     4.3E-05   &     6.9           \\           
07      &	6:48:57.26      &  	 93           	&       88.1    &       91.4    	&	-101.89496	& 	35.1039 	&	 5.5   &  	0.14       &  	 2.3       &     8.1E-05   &     6.9           \\           
08      &	6:48:57.51      &  	 93           	&       85.0    &       87.8    	&	-102.35109	& 	34.95404	&	 9.0   &  	0.45       &  	 1.9       &     7.0E-05   &     7.1           \\           
09      &	6:48:57.56      &  	 93           	&       90.2    &       92.0    	&	-102.1878 	& 	35.13408	&	 3.1   &  	0.15       &  	 3.3       &     1.2E-05   &     7.0           \\           
10      &	6:48:57.56      &  	 98           	&       90.7    &       93.7    	&	-102.19213	& 	34.97145	&	 8.3   &  	0.50       &  	 1.5       &     1.6E-04   &     6.5           \\           
11      &	6:48:57.61      &  	 96           	&       89.2    &       92.5    	&	-102.18736	& 	34.99368	&	 7.7   &  	0.43       &  	 2.1       &     1.1E-04   &     6.7           \\           
12      &	6:48:57.61      &  	 92           	&       89.1    &       90.9    	&	-102.17244	& 	34.95659	&	 3.2   &  	0.21       &  	 3.6       &     4.9E-05   &     6.7           \\           
13      &	6:48:57.71      &  	 95           	&       88.3    &       91.6    	&	-102.18328	& 	34.98407	&	 7.6   &  	0.43       &  	 2.2       &     1.1E-04   &     6.7           \\           
14      &	6:48:57.81      &  	 95           	&       92.4    &       94.6    	&	-102.00226	& 	35.19183	&	 3.0   &  	0.14       &  	 3.0       &     7.6E-05   &     6.4           \\           
15      &	6:48:58.86      &  	 99           	&       92.6    &       94.4    	&	-102.22673	& 	35.22975	&	 7.4   &  	0.29       &  	 1.5       &     1.1E-04   &     6.5           \\           
16      &	6:48:58.16      &  	 97           	&       92.5    &       94.9    	&	-101.87972	& 	35.16405	&	 5.1   &  	0.21       &  	 2.5       &     8.5E-05   &     6.5           \\           
17      &	6:48:58.30      &  	 96           	&       93.2    &       95.0    	&	-101.71504	& 	34.97457	&	 3.0   &  	0.14       &  	 3.6       &     3.2E-05   &     6.4           \\           
18      &	6:48:58.38      &  	 93           	&       89.7    &       90.5    	&	-102.15558	& 	35.00947	&	 4.0   &  	0.14       &  	 3.6       &     1.6E-05   &     6.9           \\           
19      &	6:48:58.44      &  	 92           	&       89.3    &       90.3    	&	-102.15132	& 	35.00131	&	 3.0   &  	0.15       &  	 3.8       &     5.7E-06   &     7.1           \\           
20      &	6:48:58.61      &  	 91           	&       86.7    &       90.7    	&	-102.35982	& 	34.80349	&	 4.8   &  	0.25       &  	 3.4       &     1.1E-05   &     7.2           \\           
21      &	6:48:58.71      &  	 91           	&       86.8    &       89.0    	&	-102.31085	& 	35.0673 	&	 4.8   &  	0.25       &  	 3.1       &     2.7E-05   &     7.2           \\           
23      &	6:48:58.84      &  	 94           	&       90.1    &       91.2    	&	-101.57977	& 	35.0429 	&	 4.4   &  	0.21       &  	 3.2       &     4.4E-05   &     6.7           \\           
24      &	6:48:58.86      &  	 91           	&       87.3    &       89.3    	&	-102.0335 	& 	35.18851	&	 4.2   &  	0.20       &  	 3.0       &     1.2E-05   &     7.3           \\           
26      &	6:48:59.21      &  	 93           	&       88.9    &       90.8    	&	-101.67051	& 	35.07256	&	 4.8   &  	0.21       &  	 2.9       &     5.2E-05   &     6.8           \\           
27      &	6:48:59.21      &  	 88           	&       84.7    &       87.2    	&	-101.668  	& 	35.07163	&	 3.8   &  	0.14       &  	 3.9       &     7.1E-06   &     7.4           \\           
28      &	6:48:59.36      &  	 97           	&       92.5    &       94.2    	&	-101.96086	& 	34.89468	&	 5.2   &  	0.29       &  	 2.5       &     8.9E-05   &     6.5           \\           
31      &	6:48:59.56      &  	101           	&       92.2    &       97.0    	&	-102.10948	& 	34.93176	&	10.0   &  	0.43       &  	 0.7       &     4.8E-04   &     6.3           \\           
32      &	6:48:59.56      &  	 97           	&       92.1    &       95.0    	&	-102.13669	& 	35.01049	&	 5.6   &  	0.29       &  	 2.0       &     1.7E-04   &     6.4           \\           
33      &	6:48:59.76      &  	 96           	&       89.5    &       92.8    	&	-101.94317	& 	35.18404	&	 7.3   &  	0.36       &  	 2.5       &     8.6E-05   &     6.7           \\           
34      &	6:48:59.81      &  	 95           	&       91.0    &       92.7    	&	-101.68839	& 	35.08403	&	 4.5   &  	0.14       &  	 2.5       &     7.1E-05   &     6.7           \\           
35      &	6:48:59.81      &  	101           	&       90.6    &       95.0    	&	-101.82804	& 	35.19464	&	11.8   &  	0.57       &  	 0.0       &     6.3E-04   &     6.4           \\           
36      &	6:48:59.86      &  	98              &       94.4    &       94.6        &	-101.89538	& 	35.0223 	&	7.4    &  	0.21       &  	 2.0       &     1.5E-04   &     6.5           \\           
37      &	6:49:00.01      &  	100           	&       90.2    &       93.2    	&	-101.68179	& 	34.54581	&	11.1   &  	0.64       &  	 1.3       &     2.6E-04   &     6.4           \\           
38      &	6:49:00.11      &  	 95           	&       90.3    &       93.3    	&	-101.42246	& 	35.11125	&	 5.5   &  	0.29       &  	 3.4       &     3.1E-05   &     7.0           \\           
39      &	6:49:00.51      &  	 93           	&       90.5    &       91.9    	&	-101.80418	& 	34.46058	&	 2.7   &  	0.21       &  	 3.8       &     6.3E-05   &     6.8           \\           
40      &	6:49:00.56      &  	 98           	&       92.6    &       95.6    	&	-101.60873	& 	35.04   	&	 6.1   &  	0.36       &  	 2.5       &     8.6E-05   &     6.6           \\           
41      &	6:49:00.61      &  	 99           	&       93.9    &       97.6    	&	-101.28071	& 	34.5796 	&	 7.0   &  	0.43       &  	 1.9       &     1.6E-04   &     6.3           \\           
42      &	6:49:00.71      &  	 96           	&       86.6    &       93.0    	&	-101.81478	& 	34.88312	&	10.7   &  	0.43       &  	 2.2       &     1.1E-04   &     6.7           \\           
43      &	6:49:00.71      &  	 92           	&       89.9    &       90.2    	&	-101.46805	& 	35.03655	&	 2.4   &  	0.14       &  	 3.4       &     4.3E-05   &     6.7           \\           
44      &	6:49:00.81      &  	 98           	&       92.4    &       94.6    	&	-101.93843	& 	35.16374	&	 6.4   &  	0.36       &  	 1.7       &     1.6E-04   &     6.8           \\           
45      &	6:49:01.01      &  	 95           	&       90.4    &       92.7    	&	-101.73432	& 	34.96063	&	 5.2   &  	0.29       &  	 3.0       &     2.9E-05   &     6.7           \\           
46      &	6:49:01.64      &  	 99           	&       93.4    &       96.3    	&	-101.4016 	& 	34.49924	&	 7.6   &  	0.43       &  	 2.9       &     5.7E-05   &     6.3           \\           
47      &	6:49:02.01      &  	 99           	&       90.1    &       93.4    	&	-101.68947	& 	35.07359	&	10.0   &  	0.57       &  	 1.5       &     2.1E-04   &     6.5           \\           
48      &	6:49:02.06      &  	 98           	&       90.9    &       94.4    	&	-101.49462	& 	34.9333 	&	 8.0   &  	0.50       &  	 1.1       &     3.6E-04   &     6.6           \\           
50      &	6:49:03.31      &  	 96           	&       89.6    &       92.8    	&	-101.32526	& 	35.15653	&	 7.2   &  	0.43       &  	 1.7       &     1.4E-04   &     6.7           \\           
51      &	6:49:04.31      &  	 93           	&       88.7    &       90.8    	&	-101.15794	& 	34.812  	&	 4.9   &  	0.21       &  	 2.9       &     4.6E-05   &     7.0           \\           
52      &	6:49:04.45      &  	 93           	&       89.4    &       91.8    	&	-101.49087	& 	34.62518	&	 4.0   &  	0.14       &  	 3.8       &     2.5E-05   &     6.8           \\           
\hline
\end{tabular}
\tablefoot{$met$: meteor number, $time$: time, $H_{BEG}$: beginning height, $H_{END}$: end height, $H_{MAX}$: height of maximum brightness, $\lambda_{B}$ and $\phi_{B}$: longitude and latitude of the beginning point, $l$: length of trajectory, $t_{dur}$: duration of meteor, $M_{MAX}$: maximum brightness, $m_{phot}$: photometric mass, $K_{B}$: Ceplecha's parameter.}
\end{table*}

\subsection{Physical properties and comparison with other $\tau$-Herculid meteoroids}
\label{physical}

The reconstruction of the atmospheric trajectories of the cluster meteors is based on the assumption that they all belong to the $\tau$-Herculid meteor shower. We can verify this by comparing their physical properties with meteors of known shower membership. Detailed analyses of the 2022 $\tau$-Herculid meteors have been made by \citet{Koten2023}. The 2022 $\tau$-Herculid paper is based on more than 200 meteors, 80 of which were double or multi-station. They cover a much wider range of the photometric masses of the meteors, which extends from 10~mg to about 10~kg. On the other hand, the photometric masses of the meteors in the cluster range from 5.75~mg to 4.5~g.

\subsubsection{Beginning heights, parameter $K_{B}$}
The beginning heights of the cluster meteors were estimated with an accuracy of 1~km. This is due to the setup of the single station meteor trajectory calculation method. A better precision is possible, but it could lead to an increase in the number of the possible solutions. Due to other errors that accumulate during the whole processing, we do not consider it necessary to estimate the beginning heights with higher resolution. Therefore, the beginning heights are integers. Taking into account the measurement errors, the overall accuracy of the the beginning height determination can be up to 2~km, especially in the case of very faint meteors.

Figure~\ref{fig_HB} shows that the beginning height (blue line) increases with the photometric mass. This is a common trend for meteor showers of the cometary origin. When compared with the 2022 $\tau$-Herculid meteors recorded before the peak of activity (red line) \citep{Koten2023}, the slope of this dependence is similar. Note that the Maia camera has twice the frame rate and is therefore slightly less sensitive, resulting in slightly lower meteor beginning heights. The fact that the cameras capturing the cluster were placed above the most densest part of the atmosphere may also play a role.

\begin{figure}
  \resizebox{\hsize}{!}{\includegraphics{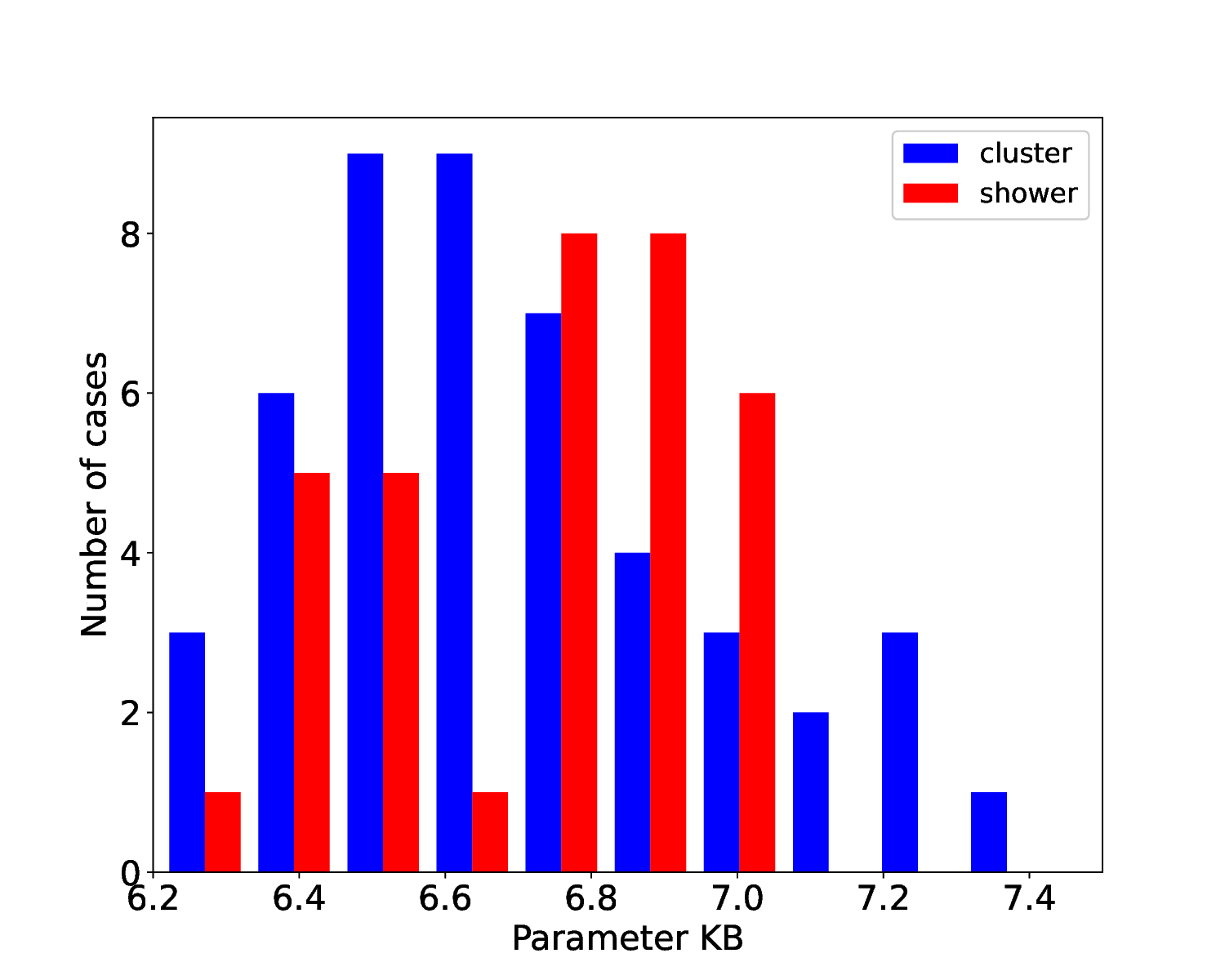}}
  \caption{Ceplecha's $K_{B}$ parameter of the cluster meteors (blue) and their comparison with 34 shower meteors (red) recorded by the Maia video camera in the Czech Republic.}
  \label{fig_KB_hist}
\end{figure}

The limiting stellar magnitude for AMOS-Spec cameras was measured to be around +5.0, meteors are first detected when they reach a magnitude of 4.3-4.5. For this reason, the value of the $K_{B}$ parameter was corrected by +0.15 \citep{Ceplecha1967}, which is the usual value for the video cameras. 

Note that the $K_{B}$ parameter corrects the beginning height for the initial velocity and slope of the trajectory. Its use therefore allows experiments to be compared regardless of camera sensitivity. It is defined by the equation
\begin{equation}
K_{B} = \log \rho_{B} + 2.5 \log v_{\infty} - 0.5 \log \cos z_{R} + 0.15,
\end{equation}
where $\rho_{B}$ is the atmospheric density [g~cm$^{-3}$] at the beginning height $H_{B}$, $v_{\infty}$ is the initial velocity [cm~s$^{-1}$], and $z_{R}$ is the zenith distance of the radiant.

According to Ceplecha's classification \citet{Cep1988}, the majority of the meteors is classified as a type C, i.e. as a regular cometary material. With respect to the orbit of the meteor shower in the Solar System, it is a subtype of C1, i.e. short-period comets. There are several out-layers belonging to the group B. Nevertheless, the mean value $K_{B} = 6.7\pm 0.3$, i.e. still the class C. A similar range of values was found for the meteor shower, with a mean value of $K_{B} = 6.9\pm 0.2$ \citep{Koten2023}. 

Figure~\ref{fig_KB_hist} illustrates a comparison of the distribution of the $K_{B}$ parameter between cluster meteors and the shower meteors observed by the Maia video camera. The sensitivity of both types of cameras is similar. Despite a small shift towards higher $K_{B}$ values in the histogram for the cluster meteors compared to the shower meteors, the majority of the meteors from both samples belong to the same class. Moreover, we should keep in mind that the beginning heights of the cluster meteors are estimated with a precision of 2 km, which could also influence the value of the $K_{B}$ parameter.

\subsubsection{Mass distribution}
\label{massDistribution}
For the pre-peak activity during the 2022 outburst, the mass distribution index was found to be $s = 2.02\pm0.12$ \citep{Koten2023}. The Global Meteor Network reported almost the same value of $s = 2.0$, a measurement which is based on a higher number of meteors \citep{Egal2023}. 

\begin{figure}
  \resizebox{\hsize}{!}{\includegraphics{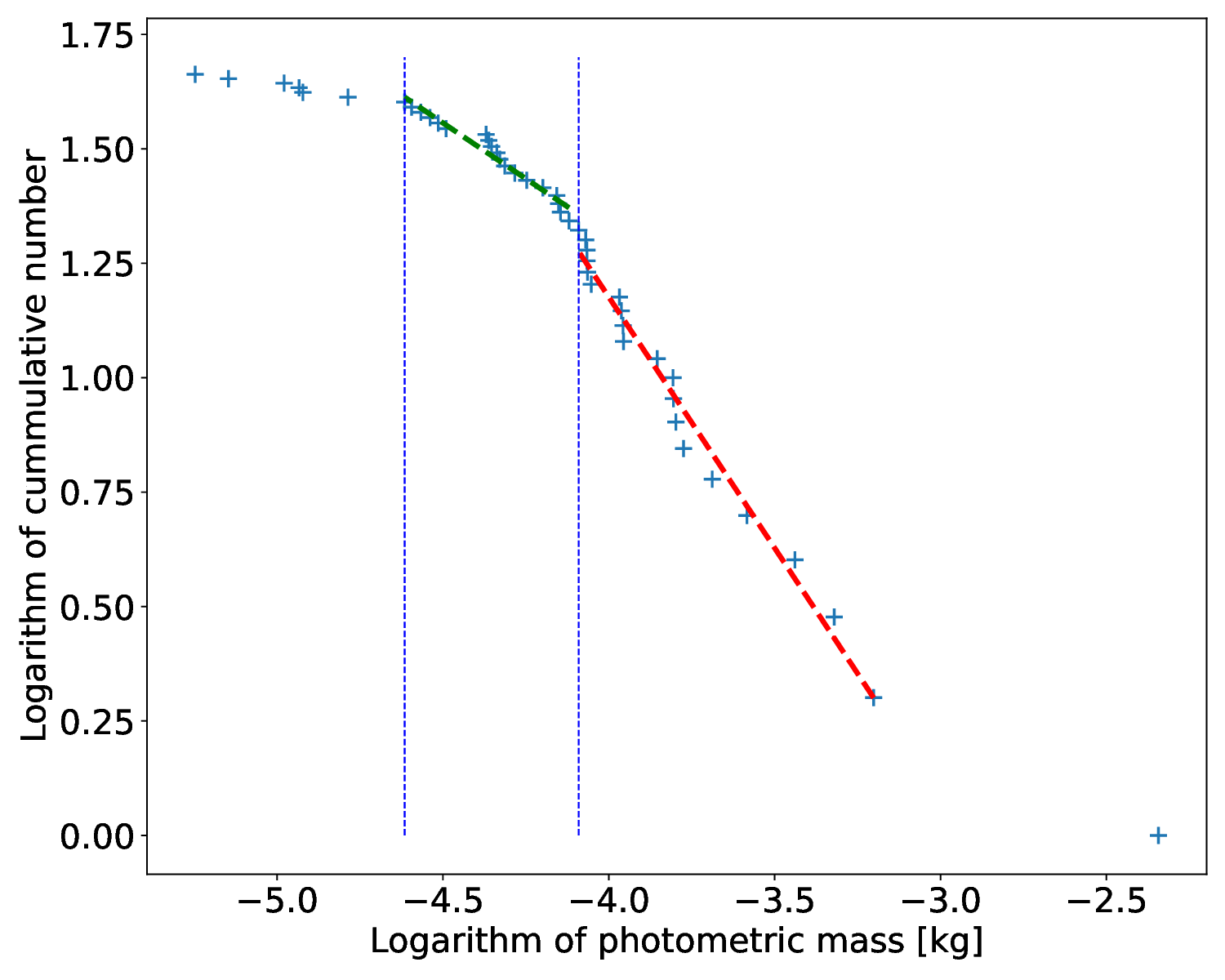}}
  \caption{Cumulative distribution of the cluster meteor masses. The red dashed line represents a linear part of the plot, which is fitted with a line of slope $k = -1.10$ and which was used for the calculation of the mass distribution index. The green dashed line fits masses between 0.02 g and 0.08 g. These limits, used to calculate the total mass of the cluster, are marked by blue dashed vertical lines. The slope of this line is $k = -0.49$.}
  \label{fig_mass_distribution}
\end{figure}

The mass distribution for the cluster members can be better described by two power-laws with different slopes, as shown in Figure~\ref{fig_mass_distribution}. For fragments with masses above $0.08$~g,the measured slope of the fit is $k = -1.1$, giving a mass index of $s=2.1$, very similar to that of the main peak value. Taking into account the uncertainty of the single station trajectories, the agreement between these cluster members and shower meteors in terms of mass distribution is good.  For fragments with masses between $0.02$~g and $0.08$~g the slope is smaller, with $s=1.49$. The photometric mass of 47 cluster members is $9.357$~g. However, the total mass of the cluster is probably higher because we can assume that we did not observe all the fainter meteors produced by the less massive fragments. We assume that our record is incomplete for fragments with masses below 0.024 g and we estimate their numbers by extrapolating a power law function with index $s=1.49$. Using the same method as \citet{Capek2022}, with values of $D=1.49$, $m_{\rm S}=m_{05}=0.024$~g, $m_{\rm L}=m_{14}=0.076$~g, we estimate the total mass of fragments smaller than $0.024$~g to be $1.13$~g. The corresponding observed fragments (there are $6$ of them) have a mass of only $0.064$~g. We estimate that the total mass of the whole cluster is $10.43$~,g. This means that the mass of the brightest fragment is 43\% of the total mass of the cluster. This is still the mass dominant fragment, although not as dominant as in the case of the previously analysed clusters.

The atmospheric trajectories of the meteors were estimated on the basis of the assumption that they belonged to the meteor shower. The calculated properties, including the mass, mass distribution index and the class of the meteoroids, were found to be very similar to those of the $\tau$-Herculid meteor shower. This supports the initial idea that the cluster indeed belongs to this shower.

\subsection{Statistical analyses of cluster appearance}
\label{statistics}

The meteor cluster consists of 52 meteors, which were observed over a period of 8.5 seconds. On the other hand, data from several sources, summarised by \citet{Egal2023}, indicate that the ZHR at the shower's maximum of activity was about 27 hr$^{-1}$. The peak of the activity occurred between 4 and 4:30 UT, so the cluster was observed 2 to 2.5 hours later. Furthermore, the activity profile \citep{Egal2023} was compiled from several data sources, including the Global Meteor Network \citep{Vida2022}, and no such an increase in activity was observed at the time of the cluster. This means that it was not the global feature of the activity profile, but the local and brief increase in activity. 

Such short bursts of the activity have only been recorded during the meteor storms of the Leonid meteor shower when the activity reaches thousands of meteors per hour \citep{Kinoshita1999, Watanabe2002, Watanabe2003}. We have therefore simulated how often such an outburst could occur during the activity of a meteor shower with a given ZHR. The random appearance of the cluster is detected if at least $52$ meteors occur within $8.5$ seconds. The meteor times are randomly generated with constant ZHR lasting 4 hours over ten million simulation runs. Clearly, such a cluster could not be observed for a shower with a ZHR $27$~hr$^{-1}$. Therefore, we increased the simulated ZHR to $52$~hr$^{-1}$. However, even with such a ZHR, the meteor cluster still did not appear in any of the ten millions of simulation runs.

We only found a non-zero probability of random appearance of such a cluster at unrealistically high activities, in which case we had to reduce the number of simulation runs to $100\,000$ due to time consumption. The results of the simulations are summarised in Table~\ref{tab_occurence_simulation}. It is obvious that a random occurrence of such a cluster, as observed during the $\tau$-Herculid shower, becomes non-negligible for a shower with a activity of about $10\,000$ meteors per hour. We can therefore exclude the possibility that the observed meteors were randomly clustered.

\begin{table}
\caption{The probability of occurrence of 52 meteors in 8.5 seconds during 4 hours for different  meteor shower activity. The probability calculation is based on the simulation of randomly generated meteor times for a meteor shower with the given ZHR. More details can be found in the text.}
\label{tab_occurence_simulation}      
\centering          
\begin{tabular}{r r} 
\hline
ZHR		& 	Probability		\\
\hline
52         &   0.0\%   \\
10\,000    &   0.4\%   \\
11\,000    &   5.3\%   \\
12\,000    &   36.6\%    \\     
13\,000    &   93.6\%    \\
\hline                  
\end{tabular}
\end{table}

\subsection{Mutual positions of fragments}
\label{positions}

Using the reconstructed atmospheric trajectories of the meteors, we can analyse their mutual positions and the overall structure of the cluster. The positions of the fragments are corrected for atmospheric deceleration before the start of ablation and for the influence of the Earth's gravitational field. A detailed description of the method to derive these corrections can be found in Appendix A of \cite{Capek2022}. 

\begin{figure*}
  \resizebox{\hsize}{!}{\includegraphics{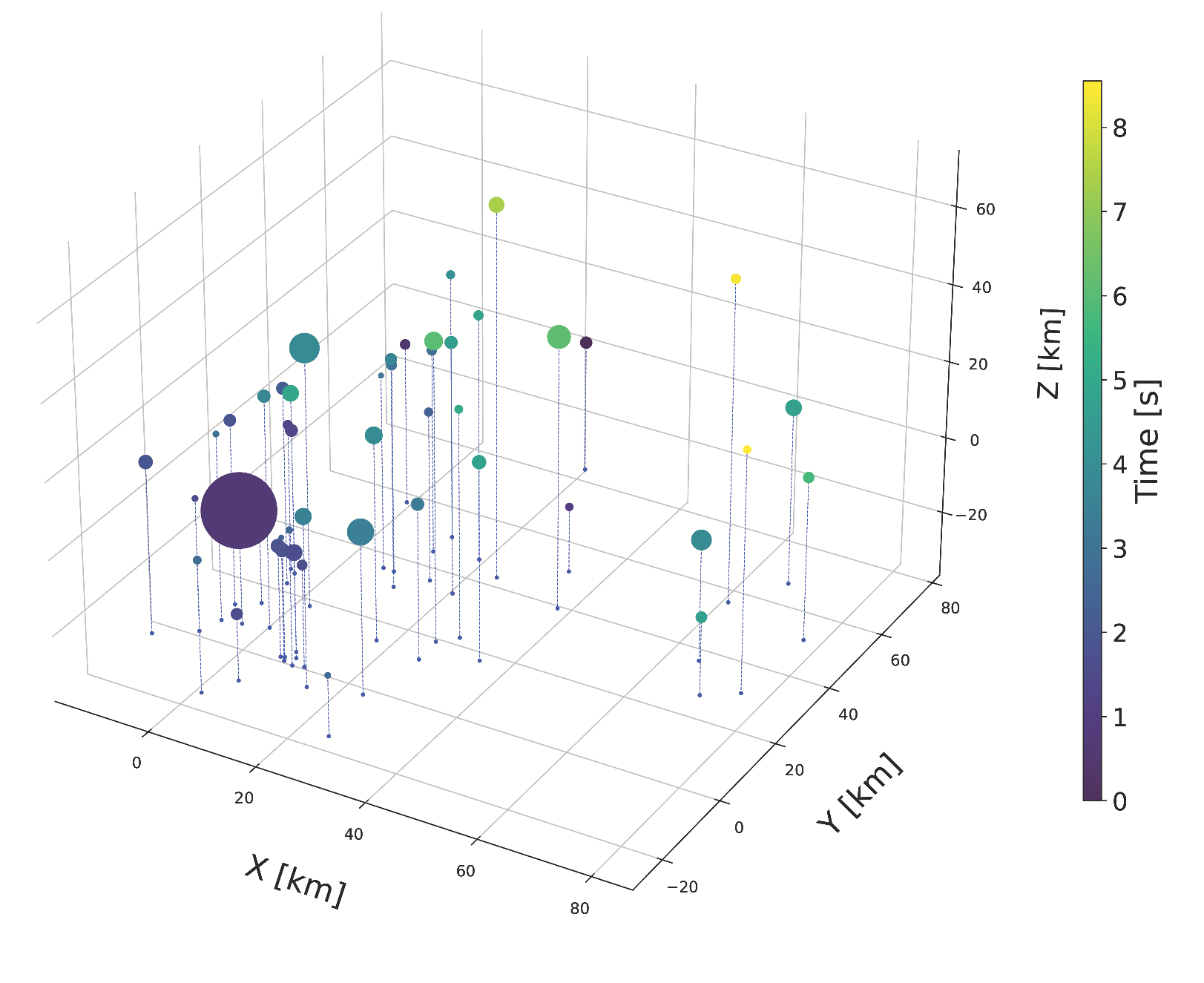}} 
  \caption{Real positions of the fragments with respect to the largest one in the coordinated system, where the x-axis is pointed in the antisolar direction and the z-axis points to the northern ecliptic pole. The size of each symbol is proportional to the mass of the fragment. Again, the beginning time of each meteor is colour-coded. For better illustration, a 'ground' projection of the fragments is marked with blue dashed lines.}
  \label{fig_positions}
\end{figure*}

The centre of the coordinate system is moved to the position of the brightest fragment. The x-axis points in the antisolar direction, the z-axis to the north pole of the ecliptic. The positions of the fragments within the cluster in this coordinate system are shown in Figure~\ref{fig_positions}, with the size of the symbols proportional to the photometric mass of each fragment. In addition, the beginning time of each meteor relative to the beginning of the first meteor is indicated by different colour of the symbol. In this representation the total volume of the cluster is $\approx90\times100\times80$~km, which is $\approx720\,000$~km$^3$.

There are two points that this plot illustrates. First, it confirms that the positions of the meteors are not perfectly aligned in time as was the case for example for the cluster detected over Scandinavia \citep{Koten2024}. This was already suggested by Figure~\ref{fig_3D_trajectories}. Second, there is no evident mass separation of the fragments. This means some smaller masses are observed closer to the main fragment and conversely some larger masses are observed further away. 

\begin{figure}
  \resizebox{\hsize}{!}{\includegraphics{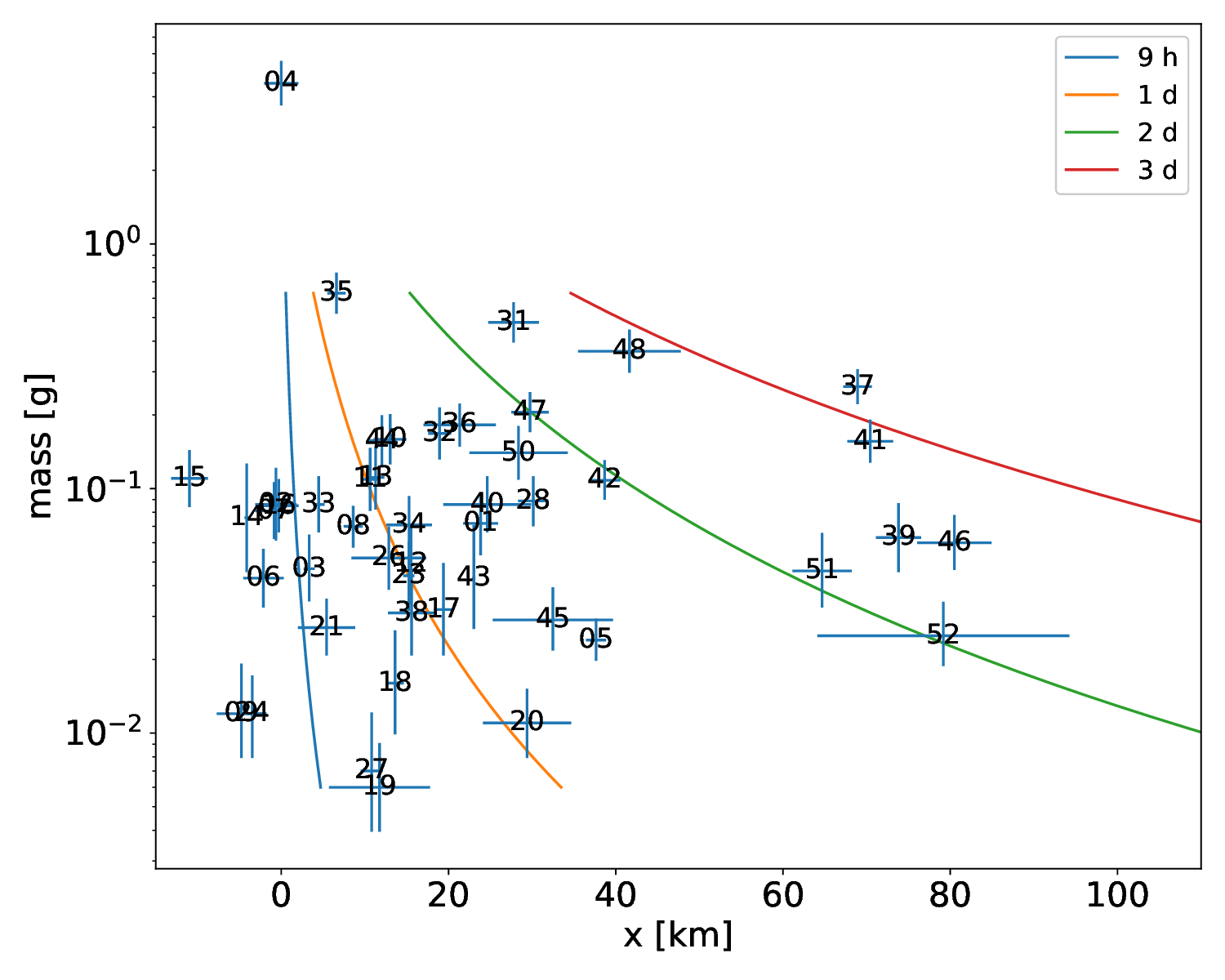}}
  \caption{Distance of each fragment from the main body as a function of its mass measured in the antisolar direction. The coloured lines represent the expected positions of the fragment in the case of zero ejection velocity. Fragments are numbered as in Table~\ref{tab_atm_traj}.}
  \label{fig_antisolar_distance}
\end{figure}

This is also confirmed by Figure~\ref{fig_antisolar_distance}, which shows the distance of each fragment from the main body projected in the anti-solar direction as a function of its mass. A negative value means that the fragment is closer to the Sun than the largest one. For illustration, the expected positions of the fragments for different times of the ejection from the main body are also shown in the figure, assuming a zero ejection velocity. The measured positions are distributed around the curves representing ages between 9 hours and 3 days. This plot therefore indicates that the age of the cluster is very small. It also means that the time was too short for the solar radiation pressure to align the fragments according to their masses.

\section{Discussion}
\label{discussion}

Most of the cluster members are shifted in the anti-solar direction with respect to the most massive fragment. However, as seen in Section \ref{positions}, the positions of fragments on the x-axis are not sorted according to mass. This fact can be characterized by Spearman's correlation coefficient of the dependence of x-axis positions on mass. Its value is $0.087$ with a p-value of $0.56$. Without using a precise statistical formulation, we can argue, because of the high p-value, that this dependence is the result of random arrangement.

The method of determining cluster ages using equation (11) in \citet{Capek2022} can be applied to cases where the solar radiation pressure has significantly affected the positions of individual fragments and the influence of ejection velocities has been overridden. This is indicated by sorting the x-positions according to the masses. For the $\tau$-Herculid cluster, this method cannot be applied. 

\begin{figure}
  \resizebox{\hsize}{!}{\includegraphics{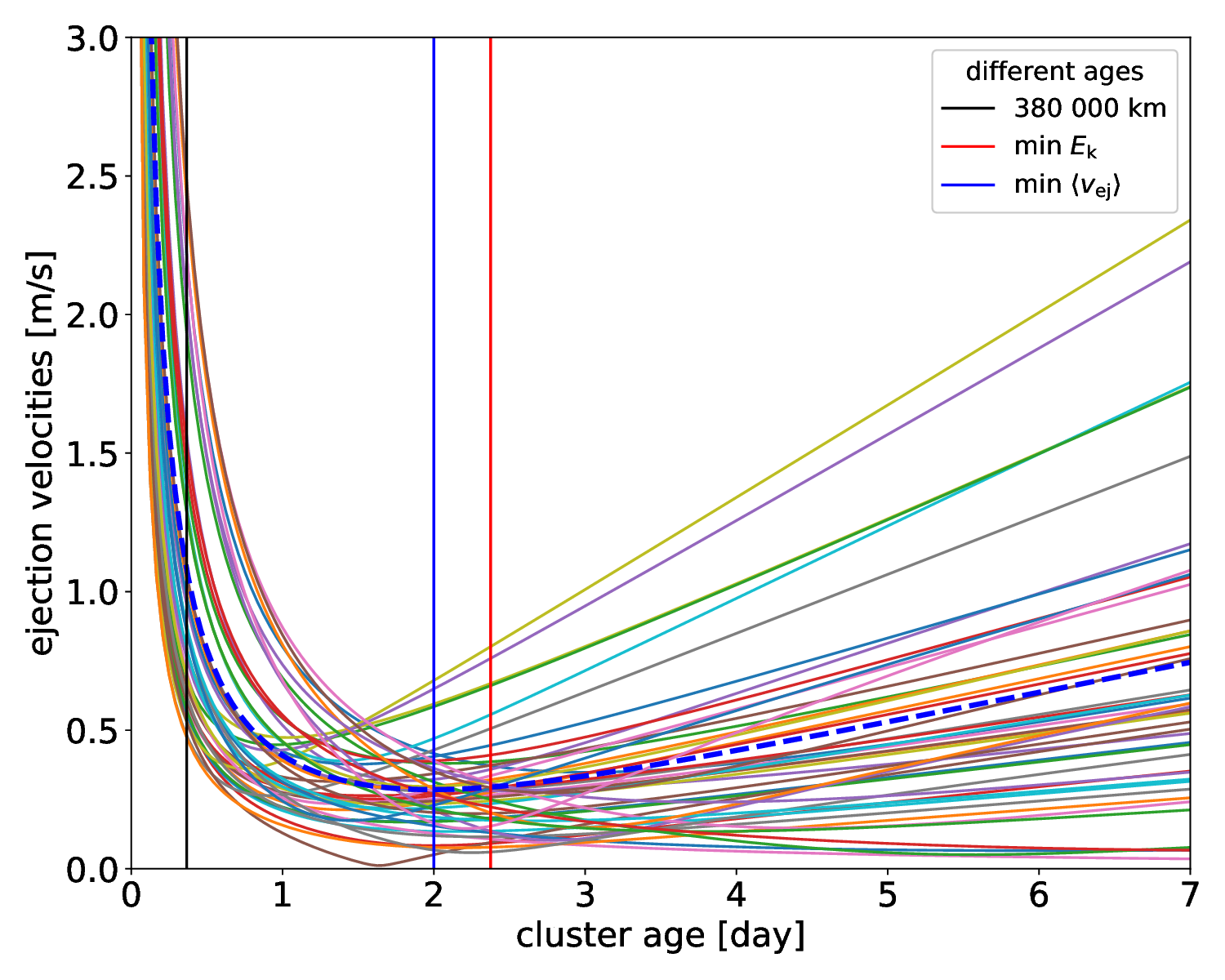}}
  \caption{Dependence of ejection velocities of individual fragments on cluster age (solid curves). The blue dashed curve corresponds to the mean ejection velocity. The vertical lines correspond to different ages: The age determined from the minimum of the initial kinetic energy of all fragments is $\sim 2.4$~days 
  (red line), while the minimum of the mean ejection velocity corresponds to an age of 
  $\sim 2.0$~days 
  (blue line). The age corresponding to the formation at the distance of the semi-major axis of the Moon's orbit ($\sim 9$~hours) is shown in black.}
  \label{fig_vej}
\end{figure}

Therefore, a different approach was chosen. For each fragment, the position relative to the cluster center of mass was determined. Then, the ejection velocity for different cluster ages was calculated to allow the fragment to reach this position, as shown in Figure~\ref{fig_vej}. For the calculation we assumed a density of the fragments of $250$~kg~m$^{-3}$. This value has been found for two modelled meteoroids with masses of less than a one gram \citep{Egal2023}. The properties of a meteoroid of at least 10 kg observed over south-eastern Europe also suggest a similar bulk density \citep{Koten2023}. 

As Figure~\ref{fig_vej} shows, the minimum of the total kinetic energy of the fragments occurs at an age of ~2.4 days. The minimum of the mean ejection velocity is reached at an age of ~2.0 days. However, we cannot take these values with confidence as the true age of the cluster.

Two remarks can be made about the ages and ejection velocities: (i) The members of the cluster are significantly more massive than in the case of the September $\epsilon$-Perseid cluster (except for the fireball), and with similar ejection velocities, the same age ($\sim 2.3$~days) would not be sufficient to sort it significantly. (ii) The ejection velocities are less than $1$~m~s$^{-1}$ in a wide range of ages; from less than $1$~day to $\sim 3$~days and for most fragments up to 7~days. For example, if the cluster formation occurred at about the distance of the Moon, corresponding to an age of less than $9$~hours, the ejection velocities would range from $0.46$~m~s$^{-1}$ to $2.42$~m~s$^{-1}$ with a mean value of $1.05$~m~s$^{-1}$.

Furthermore, we considered the possibility that the cluster was not formed during one, but during several successive fragmentations. We assumed zero $x$-components of the ejection velocities, and for each fragment we determined the time required to reach its final position due to solar radiation pressure. However, the distribution of these partial ages is random with no significant aggregations indicating successive fragmentations. We admit that such a method is too simplified, so that only a much more sophisticated model can provide an answer to the question of possible successive fragmentations, but this is beyond the scope of this study.

The formation process of the cluster cannot be reliably estimated in this case. For a wide range of possible ages, there are low ejection velocities, as shown above. In that case, similar to the September $\epsilon$-Perseid cluster and the cluster over northern Europe \citep{Capek2022, Koten2024}, the most likely cause would be disintegration due to thermal stresses. It is also supported by the fact that the $\tau$-Herculid meteoroids were found to be very fragile with extremely low bulk densities \citep{Egal2023, Koten2023}. However, a very low age of a few hours cannot be ruled out. Then, ejection velocities would significantly exceed $1$~m~s$^{-1}$, and we could also consider disintegration after the impact of a small meteoroid, or even rotational breakup. 

\section{Conclusions}
\label{conclusions}

We reanalyzed the $\tau$-Herculids meteor cluster first reported by \cite{Vaubaillon2023} after considering additional data from the AMOS camera. Additional members were spotted, for a total of 52 fragments, most of which could be fully characterized. The trajectory and relative separation shows that this cluster was young and formed shortly (<2.5 days) before its observation, preventing the Solar radiation pressure to have a significant effect on the large vs. small mass fragments relative position. Thermal stress is supposed to be the main formation process, although other scenarii cannot be excluded.

The properties of the meteors in the cluster were very similar to those of the $\tau$-Herculid meteor shower, justifying the assumption of the shower membership used to calculate the atmospheric trajectories of the single station meteors. It also expanded our set of the $\tau$-Herculid meteors available for further analysis. 

Statistical analysis confirms that the cluster cannot be a random occurrence of the meteors, as such a random grouping is only possible for the shower with an activity of more than ten thousand meteors per hour. This was not the case of the $\tau$-Herculid 2022 outburst. Therefore, the observed group of meteors was a real meteor cluster, the most numerous ever analysed in detail.

\begin{acknowledgements}
This work was supported by the Grant Agency of the Czech Republic grants 20-10907S (cluster data processing and analysis) and 24-10143S (software update and meteor shower properties), and the institutional project RVO:67985815 (institutional infrastructure). Airborne campaign organized and funded through RTI and supported by University of Southern Queensland. Dr. Fabian Zander is funded by the Australian Research Council through the DECRA number DE200101674. Dr. Juraj Toth was supported by ESA contract No. 4000128930 /19/NL/SC, the Slovak Research and Development Agency grants APVV-16-0148 and APVV-23-0323, the Slovak Grant Agency for Science VEGA 1/0218/22.
\end{acknowledgements}

\bibliographystyle{aa}
\bibliography{00pkbib}

\begin{appendix}
\label{ang_vel_note}

\section{Note on the angular velocity determination}

To demonstrate the sensitivity of the atmospheric trajectory determination to the angular velocity of the meteor, we examine the relative position of fragment \#52 to fragment \#4, the most massive. Meteor \#52 was measured in only five frames. The measured angular velocity was 7.7$^{\circ}$/s. The corresponding beginning height was 93~km. This is about 0.5~km lower than the value calculated from the Figure~\ref{fig_HB}. Note, that such a result is satisfactory. The resulting distance between the beginning points of the meteors is 81.1~km.

We artificially change the angular velocity of the meteor \#52 and see how this affects the distance between the beginning points of two meteors. The results are summarised in Table~\ref{tab_ang_vel}.

\begin{table}
\caption{The distance between the beginning points of the meteors \#4 and \#52 as a function of the angular velocity of meteor \#52.}             
\label{tab_ang_vel}      
\centering          
\begin{tabular}{r r r} 
\hline
$\omega$    	& 	$H_{BEG}$       &   $D$		   \\
$[^{\circ}/\rm{s}]$    &   [km]            &   [km]       \\
\hline
6.0             &   104             &   80.4       \\
7.0             &   100             &   80.3       \\
7.7             &   93              &   81.1       \\
8.5             &   81              &   83.0       \\     
\hline                  
\end{tabular}
\tablefoot{$\omega$: angular velocity of meteor, $H_{BEG}$: beginning height, $D$: distance between meteors.}

\end{table}

It appears that the change in angular velocity does not significantly affect the distance between the beginning points of the meteors. On the other hand, for some of its values we have obtained unrealistic beginning heights of the meteors. This therefore justifies checking the correct beginning height using the Figure~\ref{fig_HB}, especially for the faintest meteors.

\end{appendix}

\end{document}